\documentstyle[psfig]{mn}

\newif\ifAMStwofonts

\newcommand{\lapp}{\mbox{\raisebox{-0.3em}{$\stackrel{\textstyle <}{\sim}$}}}
\newcommand{\gapp}{\mbox{\raisebox{-0.3em}{$\stackrel{\textstyle >}{\sim}$}}}

\title{J1432+158: the most distant giant quasar}
\author[A.K. Singal, C. Konar and D.J. Saikia]
       {Ashok K. Singal$^1$, C. Konar$^2$ and D.J. Saikia$^2$ \\
$^1$ Astronomy and Astrophysics Division, Physical Research Laboratory, 
Navrangpura, Ahmedabad 380 009, India \\
$^2$ National Centre for Radio Astrophysics, TIFR, Pune University Campus, Post Bag 3, 
Pune 411 007, India}
\date{Accepted.    Received }

\pagerange{\pageref{firstpage}--\pageref{lastpage}}
\pubyear{  }

\begin{document}

\maketitle

\label{firstpage}

\begin{abstract}
We present low-frequency, GMRT (Giant Metrewave Radio Telescope)
observations at 333 and 617 MHz of the most-distant giant quasar, J1432+158,
which is at a redshift of 1.005. The radio source 
has a total angular extent of 168 
arcsec, corresponding to a projected linear size of 1.35 Mpc. This makes it
presently the largest single object observed beyond a redshift of one.
The objectives of the GMRT observations
were to investigate the possibility of detecting a bridge of emission at low frequencies,
which may be suppressed due to inverse-Compton losses against the cosmic microwave 
background radiation.  We detect a jet-like structure connecting the core to the western
hotspot, while the eastern hotspot is found to be largely tail-less with no significant
bridge emission. The estimated life-time for the radiating electrons in the tail of the western lobe
appears smaller than the travel time of the radiating particles from the hotspot, suggesting either in-situ
acceleration or dissipation of energy by the jet at this location. 
The pressure of the intergalactic medium at $z\sim1$ estimated from the minimum energy density calculations 
appears to be marginally lower than the value extrapolated from nearby giant radio galaxies.
\end{abstract}


\section{Introduction}

The giant radio sources (GRSs), defined to be those with a projected linear 
size greater than about a Mpc (H$_o$=71 km s$^{-1}$ Mpc$^{-1}$, $\Omega_m$=0.27,
$\Omega_{vac}$=0.73, Spergel et al. 2003), 
are the largest single objects in the Universe, extending well beyond the
haloes of their parent galaxies. The GRSs are useful for studying
the evolution of radio sources with age, and also probing the intergalactic medium (IGM)
and its evolution with redshift. Statistical analysis of known giants 
show a dearth of GRSs of high luminosity, 
and no objects extending beyond about 2 Mpc, with the sole exception of 3C236 
(e.g. Ishwara-Chandra \& Saikia 1999, hereinafter referred to as IC99; 
Schoenmakers et al. 2000, 2001, and references therein).  
Although the distribution of radio sources in the luminosity-linear size or 
P-D diagram is broadly consistent with models for the evolution of radio sources
(e.g. Kaiser \& Alexander 1999; Blundell et al. 1999 and references therein),
it is not clear whether the sharp drop in the number of objects 
beyond about 2 Mpc is also partly due to possible selection effects. 

\begin{figure*}
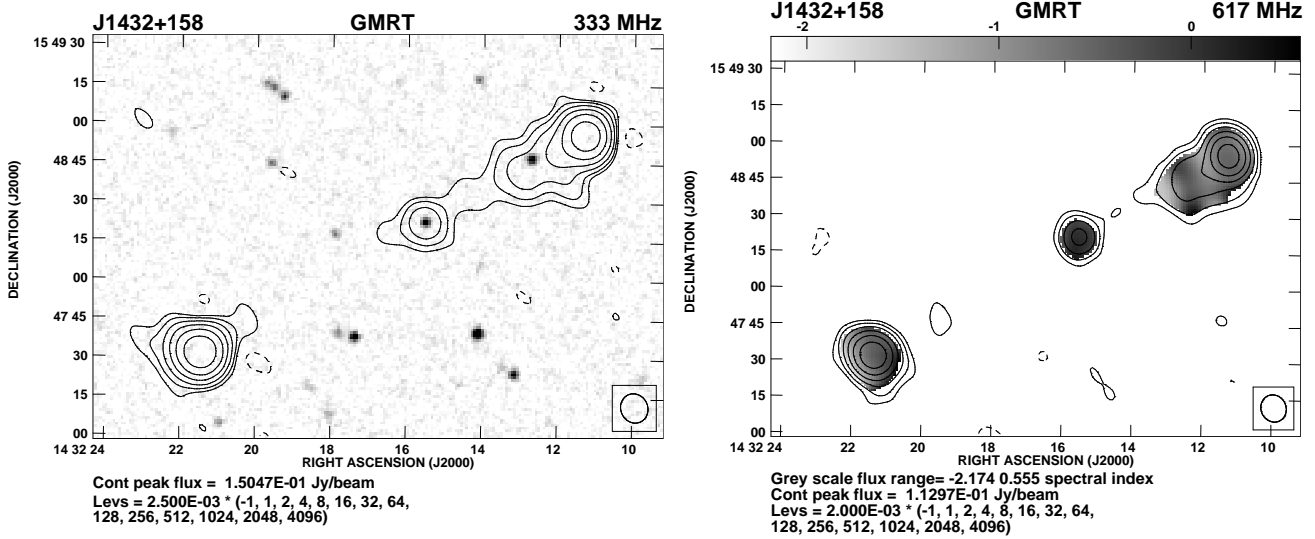

\hbox{
  \psfig{file=J1432+158_325Gn.ps,width=3.5in,angle=-90}
  \psfig{file=J1432+158_610Gn.ps,width=3.3in,angle=-90}
   }
\caption[]{Left panel:
The 333-MHz image
with an angular resolution of 11.$^{\prime\prime}$4$\times$10.$^{\prime\prime}$3 along a position
angle of 24$^\circ$ is shown superimposed on the optical field.
Right panel: The 617-MHz image is shown convolved to the same resolution as the 333-MHz
image with the spectral index between 333 and 617 MHz being shown in gray.
}
\end{figure*}

There is also a dearth of GRSs at cosmologically interesting redshifts of 
greater than $\sim$1. In the WENSS sample, the median redshift of the new GRSs
is only 0.1404 compared with 0.099 for the already known GRSs in the sample. The marginally
higher redshift is due to the WENSS survey selecting sources at lower flux 
densities (Schoenmakers et al. 2001). A number of
GRSs selected from the 7C survey have redshifts in the range of 0.3 to 0.9
(Cotter, Rawlings \& Saunders 1996). Two of the giants from the MRC 1-Jy sample,
which were observed by IC99, namely 0437$-$224 and 1025$-$229 have redshifts of   
0.84 and 0.309 respectively. For a new sample of large radio sources compiled from the
NVSS and FIRST surveys (Machalski, Jamrozy \& Zola 2001; Chy\.{z}y et al. 2003), 
the redshifts range from $\sim$0.07 to
0.8 with a median value of $\sim$0.2. The highest redshift giant source known so
far is 4C 39.24, which is associated with a galaxy at a redshift of 1.88 and has been 
studied in some detail by Law-Green et al. (1995).

In this paper, we present low-frequency, GMRT (Giant Metrewave Radio Telescope) 
observations at 333 and 617 MHz of the most-distant giant quasar, J1432+158,
which is at a redshift of 1.005 (see Hintzen, Ulvestad \& Owen 1983), 
so that 1$^{\prime\prime}$ corresponds to 8.051 kpc,
and discuss the nature of this source. One of the effects
which makes it difficult to identify giants at large redshifts is the suppression of
bridge emission by inverse-Compton losses against the cosmic microwave background radiation,
which increases strongly with redshift. This could lead to `tail-less' hotspots leading to
their classification as independent radio sources (e.g. Baldwin 1982). The problem 
would be more acute if there
are no detected cores and radio jets. The objectives of the GMRT observations
were to investigate the possibility of detecting bridges of emission at low frequencies,
 and/or a radio jet, with the broader goal of developing strategies for identifying 
high-redshift giant sources. 

\section{Observations}

The GMRT observations were made at 333 and 617 MHz on 2002 Aug 24 and 2003 Jul 11 respectively.
The GMRT consists of thirty 45-m antennas in an approximate `Y'
shape similar to the Very Large Array but
with each antenna in a fixed position. 
Twelve antennas are randomly placed within a central
1 km by 1 km square (the ``Central Square'') and the
remainder form the irregularly shaped Y (6 on each arm) over a total extent of about 25 km.  
Further details about the array can be found at the GMRT website
at {\tt http://www.gmrt.ncra.tifr.res.in}. 
The observations were made in the standard fashion, with
each source observation interspersed with observations
of the phase calibrator. The source was observed for about 8 hours at 333 MHz, 
although a significant amount of data had to be edited out due to 
ionospheric disturbances.  The observations at 610 MHz were made in the `snapshot' mode
with only about one hour being spent on the source. The primary flux density calibrator was
3C286 whose flux density was estimated on the VLA scale. The phase calibrator at 333 MHz
was 3C298 whose flux density was estimated to be 29.6$\pm$0.3 Jy, while at 610 MHz the 
phase calibrator was J1445+099 with a flux density of 2.39$\pm$0.02 Jy.   The data analyses
were done using the Astronomical Image Processing Software (AIPS) of
the National Radio Astronomy Observatory.

\section{Results}

The GMRT images of J1432+158 at 333 and 617 MHz are presented in Fig. 1. The 333-MHz image
with an angular resolution of 11.$^{\prime\prime}$4$\times$10.$^{\prime\prime}$3 along a position 
angle of 24$^\circ$ is superimposed on the optical field, showing the radio core to be coincident
with the optical quasar. The 617-MHz image is shown convolved to the same resolution as the 333-MHz
image with the spectral index image, $\alpha$, (defined as S$\propto\nu^\alpha$), between 333 and 617 MHz
being shown in gray.
Both the images show the three main components of the source, the western lobe, the weak nuclear 
component coincident with the position of the optical quasar, and the eastern lobe. The 333-MHz
image shows evidence of a jet-like extension towards the north-western component, but no bridge
of emission is seen between the eastern hot-spot and the radio core. While the western hotspot appears
to have a tail of emission extending towards the core, no significant tail is seen in the eastern
hotspot at either 333 or 617 MHz. These images are consistent with the NVSS and FIRST images which are 
shown in Fig. 2. The higher-resolution FIRST image shows the core to be extended, with a deconvolved
angular size of 1.$^{\prime\prime}$6$\times$1.$^{\prime\prime}$0 along a position
angle of 117$^\circ$, consistent with the direction of the jet-like structure seen in the 333-MHz image.

\begin{figure*}
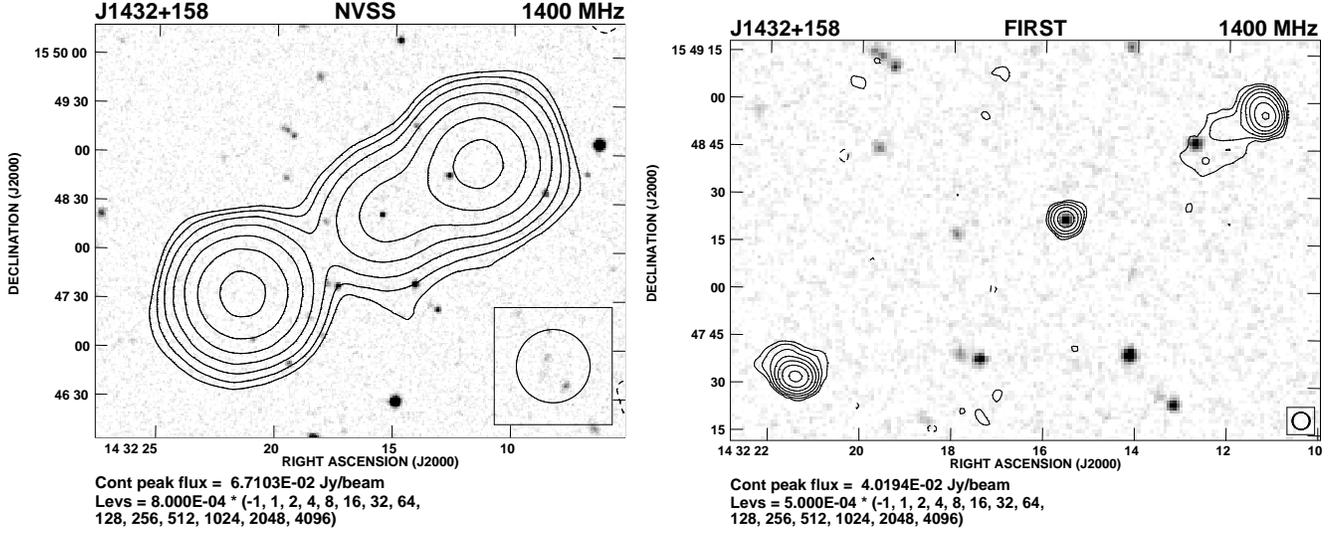

\hbox{
  \psfig{file=J1432+158_NVSSn.ps,width=3.3in,angle=-90}
  \psfig{file=J1432+158_FIRSTn.ps,width=3.6in,angle=-90}
   }
\caption[]{
The NVSS (left panel) and FIRST (right panel) images at 1400 MHz with angular resolutions
of 45 and 5.4 arcsec respectively shown superimposed on the optical field.
}
\end{figure*}

The observed parameters of J1432+158, from both the GMRT images as well as from the NVSS and FIRST images
at 1400 MHz  are listed in Table 1, which is arranged as follows.
Column 1: Name of the telescope, column 2: frequency  of observations in units of MHz,
columns 3-5: the major and minor axes of the restoring beam in arcsec and its PA in degrees;
column 6: the rms noise in units of mJy/beam; column 7: the integrated flux density of the
source in mJy estimated by specifying an area around the source; columns 8, 11 and 14: component designation;
columns 9 and 10, 12 and 13, and 15 and 16: the peak and total flux densities of each of the 
components in units of mJy/beam and mJy. 

\begin{table*}
{\bf Table 1.} The observational parameters and observed properties of J1432+158 \\

\begin{tabular}{l l l rrr r r rr r l r r r r}
\hline
Telescope  & Freq.       & \multicolumn{3}{c}{Beam size}                    & rms      & S$_I$   & Cp  & S$_p$  & S$_t$  & Cp   & S$_p$ & S$_t$ & Cp  & S$_p$   & S$_t$     \\

           & MHz         & $^{\prime\prime}$ & $^{\prime\prime}$ & $^\circ$ &    mJy   & mJy     &     & mJy    & mJy    &      & mJy   & mJy   &     & mJy     & mJy       \\ 
           &             &                   &                   &          &  /b      &         &     & /b     &        &      & /b    &       &     & /b      &           \\ \hline
GMRT       &  333        & 11.4    & 10.3       &  24                       &   0.58   &  510    & W   & 135    & 257    & C    &  20   &  26   & E   &  150    & 218       \\
GMRT       &  617        & 11.4    & 10.3       &  24                       &   0.58   &  340    & W   &  97    & 167    & C    &  20   &  22   & E   & 113     & 146       \\

VLA-NVSS   & 1400        & 45.0    & 45.0       &   0                       &   0.28   &  167    & W   &  67    &  75    & C    &  19   &       & E   &  64     &  71       \\
         
VLA-FIRST  & 1400        & 5.4     & 5.4        &   0                       &   0.15   &  142    & W   &  35    &  68    & C    &  14   & 15    & E   &  40     &  65       \\
\hline

\end{tabular}
\end{table*}

\section{Discussion and remarks}

The overall angular size of J1432+158 is 168 arcsec which corresponds to a projected
linear size of 1.35 Mpc. This makes it the largest single object 
observed beyond a redshift of one. Since it is associated with a quasar, it is expected
to be inclined at less than $\sim$45$^\circ$ to the line of sight, implying that its
intrinsic size is at least 1.9 Mpc.  Given its large size it may be relevant to 
enquire whether all the three components are physically related. Both  the outer
components have a steep radio spectrum, while the core has a flat spectrum, similar
to standard double-lobed radio sources. 
The structure of the western lobe and the presence of a 
jet-like extension from the core towards this lobe, confirms that 
these two features are indeed related. If the eastern component is unrelated, these two features
would form a weak-cored, one-sided source, which is indeed very rare (cf. Saikia et al. 1989).
Also, no optical object is found associated with the eastern component in the DSS prints.
The collinearity of the structure formed by the three main components, where the
complement of the angle formed at the nucleus by the outer hot-spots is only 2$^\circ$,
and the rough symmetry of the flux density and the location of the outer hotspots strongly 
suggest that all three components are related. The arm-length and flux density ratio of 
the lobes are $\sim$1.4 and 1.1 respectively, which is similar to that of other giant radio sources
(see IC99). However, it is the jet-side which appears to be closer to the radio core,
demonstrating the existence of asymmetries on Mpc scales either due to clusters of galaxies
or filamentary structures likely to form in hierarchical structure-formation scenarios.

\begin{figure}
\hbox{
  \psfig{file=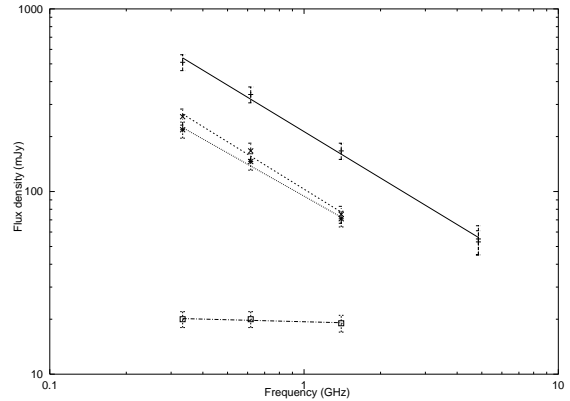,width=3.0in,angle=-90}
   }
\caption[]{
The integrated spectrum ($+$, continuous line) of J1432+158, and the spectra of the 
western ($\times$, dashed),
eastern ($\ast$, dotted) and core ($\Box$, dashed-dotted) components.
The lines show the linear least-square fits to the points. 
}
\end{figure}

The overall spectral index of the source, and the flux density of the different  components
using our measurements and values estimated from the NVSS image are shown in Fig. 3. 
The integrated flux densities at 5 GHz from the NRAO Green Bank survey are also shown
(Gregory \& Condon 1991; Becker, White \& Edwards 1991).
The spectral index of the entire source is $-$0.84$\pm$0.03, while the 
western and eastern components have steep spectral indices of $-$0.86$\pm$0.08 
and $-$0.79$\pm$0.06 respectively, and the core has a flat spectral index of $-$0.04$\pm$0.02.
The spectral index image between 333 and 617 MHz (Fig. 1) shows that the component in the
tail of the western lobe does not show evidence of any significant spectral steepening. Its
spectral index of $\sim -$0.6 between 333 and 617 MHz is similar to that of the hotspot, the
low-frequency spectrum appearing marginally flatter than the higher-frequency value. 
The flux density of the core 
measured by Hintzen, Ulvestad \& Owen (1983)  at 1415 MHz with an angular resolution of 
1.$^{\prime\prime}$6$\times$1.$^{\prime\prime}$0 is 11 mJy compared with the FIRST value of 15 mJy
with an angular resolution of 5.$^{\prime\prime}$4 arcsec. 
The difference is possibly largely due to the different resolutions of these observations.
 
The minimum energy density for a synchrotron radio source component
is given by (see e.g., Miley 1980),
\begin{eqnarray}
u_{\rm m}&= 4.5\times 10^{-11} \left[ \frac {(1+z)^{3-\alpha}
\: S_{0}\:  \nu_{0} ^{-\alpha}\: (\nu_{2} ^{0.5+\alpha} -\nu_{1} 
^{0.5+\alpha})}  {\theta_{x} \theta_{y}\: s\: (0.5+\alpha)} \right] 
^{4/7} \nonumber \\
& {\rm J\: m}^{-3}. 
\end{eqnarray}
Here $z$ is the redshift, $S_{0}$ (Jy) is the flux density at frequency 
$\nu_{0}$ (GHz), $\alpha$ is the spectral 
index with $\nu_{1}$ and $\nu_{2}$ (GHz) as the lower and upper cut off
frequencies presumed for the radio spectrum, with  
$\nu_{1} < \nu < \nu_{2}$, 
$\theta_{x}$ and $\theta_{y}$ (arcsec) represent the size 
of the source component along its major and minor radio axes, 
and $s$ is the pathlength through 
the component along the line of sight in kpc. We have assumed an equal energy 
distribution between the electrons and the heavy particles, taken 
the volume filling factor to be unity and also assumed that the
magnetic field vector lies in the plane of the sky.

The values of the minimum 
energy density and magnetic field are $3.6 \times 10^{-12}$ J m$^{-3}$ and 2.0 nT for the western hotspot and
$2.6 \times 10^{-12}$  J m$^{-3}$ and 1.7 nT for the eastern one. Here we have integrated the radio emission from
10 MHz to 100 GHz, and used the deconvolved sizes of the components.  The values for the component in the 
the tail of the western lobe are $3 \times 10^{-13}$  J m$^{-3}$ and 0.6 nT  respectively. Assuming that
the spectral break lies beyond 5 GHz,
the radiative life-time of the electron is only about $3 \times 10^{6}$  yr,
with the inverse-Compton losses being a factor of about 5 higher than the
synchrotron losses. The projected distance of this
feature from the hotspot is $\sim$220 kpc, which would take $\gapp$7$\times$10$^6$ yr for particles moving backwards with
velocities $\lapp$0.1c (e.g. Ishwara-Chandra et al. 2001).
This suggests that there could be re-acceleration of particles in this component, or it could
be part of the jet either forming a knot of emission or another hotspot. This is consistent with
the lack of spectral steepening in this region.

The minimum energy of the bridges and lobes of GRSs may be used  
to estimate the pressure in the IGM, whose emissivity is otherwise not directly detectable, 
by assuming that the lobes are in pressure equilibrium with it (e.g. Subrahmanyan and Saripalli 1993, 
hereinafter referred to as SS93; Mack et al. 1998; IC99; Schoenmakers et al. 2000). The pressure of 
such a hot, diffuse, non-relativistic IGM should vary with redshift  as 
$P_{z} \sim P_{0} (1+z)^{5} $.  SS93 estimated $P_{0}\sim10^{-15}$ Pa, based on 
Saripalli's (1988) values for minimum energy in the lobes of 8 GRSs, which appeared systematically 
lower than other similar estimates in the literature. 
Saripalli~(1988) and SS93 suggested that their 
values were lower as these represented $u_{m}$ in diffuse bridges as compared to the estimates 
made for the entire lobes in the literature. 
However, the expression for $u_{\rm m}$ used by Saripalli (1988) is lower by a factor of $\sim$3.5, 
and when corrected for this, the median value is $\sim 2\times 10^{-14}$ J m$^{-3}$,
in good agreement with other estimates in the literature. 
This corresponds to $P_{0}$ $\sim 5 \times 10^{-15}$ Pa. 
There could be a marginal trend for an increase in $P$ with redshift
(SS93; IC99; Schoenmakers et al. 2000); however, possible caveats using the 
presently available samples have been highlighted by IC99 and Schoenmakers et al. (2000). 
Our estimate for the component in the tail of the western 
lobe in the quasar J1432+158 yields $P \sim 10^{-13}$ Pa, 
while the expected value for $P_{z}$ at $z$=1.005  is $\sim 1.5\times10^{-13}$ Pa. Although
this is higher than the measured value by $\sim$1.5, estimates of $P_{z}$ in the
diffuse bridges and lobes of a sample of GRSs at cosmologically interesting redshifts of $\gapp$1,
are required to examine critically the assumption of
pressure balance with the IGM and the variation of pressure with redshift.

\section*{Acknowledgments}

The GMRT is a national facility operated by the National Centre for Radio Astrophysics
of the Tata Institute of Fundamental Research.
This research has made use of the NASA/IPAC extragalactic database (NED)
which is operated by the Jet Propulsion Laboratory, Caltech, under contract
with the National Aeronautics and Space Administration.

\end{document}